\documentclass[showpacs,amsmath,amssymb,aps,prl,twocolumn,floatfix]{revtex4-1}

\usepackage{graphicx}
\usepackage{float}
\usepackage{parskip}
\usepackage{bm}

\begin{document}

\title{Model of a $\cal PT$ symmetric Bose-Einstein condensate
  in a delta-functions double well}

\author{Holger Cartarius}
\author{G\"unter Wunner}

\affiliation{
  Institut f\"ur Theoretische Physik, Universit\"at Stuttgart,
  Pfaffenwaldring 57, 70\,569 Stuttgart, Germany}

\date{\today}

\begin{abstract}
  The observation of $\cal PT$ symmetry in a coupled optical wave guide system
  that involves a complex refractive index has been demonstrated 
  impressively in the experiment by R\"uter el al. ({\em Nat. Phys.} 6, {\bf 192}, 2010). This is, however, only an optical {\em analogue} of a quantum system, and
  it would be highly desirable to observe  the manifestation of $\cal PT$ 
  symmetry and the resulting properties also in a real, experimentally accessible, {\em quantum} system. Following a suggestion by Klaiman et al. ({\em Phys. Rev. 
    Lett.} {\bf 101}, 080402, 2008), we investigate a $\cal PT$ symmetric arrangement of a 
  Bose-Einstein condensate in a double well potential, where
  in one well cold atoms are injected while in the other particles are
  extracted from the condensate. We investigate, in particular, the effects
  of the nonlinearity in the Gross-Pitaevskii equation on the $\cal PT$ 
  properties of the condensate.  To study 
  these effects we analyze a simple one-dimensional model system 
  in which the condensate is placed into two $\cal PT$ symmetric $\delta$-function traps. 
  The analysis will serve as a useful guide for
  studies of the behaviour of Bose-Einstein condensates in
  realistic $\cal PT$ symmetric double wells. 
\end{abstract}


\maketitle

Beginning with the seminal paper by Bender and Boettcher in 1999 
\cite{bender98},
parity-time ($\cal PT$) symmetric quantum mechanics has attracted
ever increasing attention over the past decade because it offers a class of complex
Hamiltonians which, in spite of their non-Hermiticity, possess
discrete real energy eigenvalue spectra. Moreover, these Hamiltonians
feature the property of branch points, i.e., the coalescence of both 
energy values and eigenfunctions when some parameter in the Hamiltonian is varied, 
a phenomenon impossible in Hermitian quantum mechanics (but known to appear
for resonances in the continuous spectrum, see, e.g., \cite{rotter10}). 

Recently $\cal PT$ symmetry
has been realized experimentally in structured optical waveguides \cite{guo09,
  rueter10},
where the complex index of refraction is manipulated by introducing
loss and gain terms. These experiments make use of the quantum-optical 
analogy that  the wave equation for the transverse electric
field mode is formally equivalent to the one-dimensional Schr\"odinger equation.
It would, however, be
desirable to observe $\cal PT$ symmetry also in a real quantum
system. 

Klaiman et al. \cite{klaiman08a} have suggested a quantum scenario analogous
to the waveguide experiments in which a Bose-Einstein condensate
is placed in a double well potential, and loss and gain is realized by
removing atoms in one well and coherently adding particles in the
other. These authors pointed out that to have a close analogy with
the optics experiments the nonlinearity in the Gross-Pitaevskii
equation governing these condensates should be kept small. Here we want
to ask the opposite question: What are the effects of the
nonlinearity on the $\cal PT$ symmetry on such an arrangement
of a Bose-Einstein condensate? It is namely exactly the nonlinearity, 
proportional to $|\psi(x)|^2$, in the Gross-Pitaevskii equation that 
complicates matters. A necessary condition for the Hamiltonian to be 
$\cal PT$ symmetric is that the imaginary part of the potential is an 
odd function, and the real part an even function of $x$. The latter
cannot be assumed from the outset for  $|\psi(x)|^2$ when solving
the Gross-Piatevskii equation.

In this paper we will investigate the effects of the nonlinearity
on the $\cal PT$ symmetry in the spirit of a model calculation by considering the  
situation where the double well is idealized by two 
delta-function traps, with loss added in one trap and gain in the 
other. We will demonstrate that the stationary solutions of the
Gross-Pitaevskii equation indeed preserve the $\cal PT$ symmetry
of the nonlinear Hamiltonian, and merge in a branch point at some 
critical value of the 
loss and gain, beyond which the symmetry is broken. Our results show
that it will be a worthwhile enterprise to investigate $\cal PT$ symmetric
Bose-Einstein condensates in realistic double well potentials, and 
possibly pin down physical parameters where $\cal PT$ breaking
could be observed in a real experiment.

A model which mimics the physical situation of a BEC in a symmetric 
double well with loss and gain
has already been investigated by Graefe et al. 
\cite{graefe08a,graefe08b,graefe10} in the framework of 
a two-mode Bose-Hubbard-type $\cal PT$ symmetric Hamiltonian.
As an optical analogue, in the  two-mode approximation Ramezani et al. 
\cite{ramezani10} have recently looked at a mathematical model of a 
$\cal PT$ symmetric coupled dual waveguide arrangement with Kerr nonlinearity. 
It is one objective of this paper to see which features of these models
are recovered when actually solving the nonlinear $\cal PT$ symmetric 
Gross-Pitaevskii equation.

For a system where a real delta-function potential is 
augmented by a $\cal PT$ symmetric pair of delta-functions
with imaginary coefficients, bound states 
and scattering wave functions have been calculated by Jones \cite{jones08}.
His interest was devoted to the quasi-Hermitian
analysis of the problem, and no nonlinearity was present. Jakubsk\'y and Znojil \cite{jakubsky05} have considered
the explicitly solvable model of a particle exposed to two imaginary $\cal PT$ 
delta-function potentials in an infinitely high square well, and determined 
the energy spectrum.
The nonlinear Schr\"odinger
equation for a delta-functions comb was studied by Witthaut et al. \cite{witthaut08}
with the aim of gaining insight into the properties of nonlinear stationary 
states of periodic potentials.
Also, there exists a vast amount of
literature on solitons and Bose-Einstein condensates in periodic 
optical and nonlinear lattices with $\cal PT$ symmetry 
and their nonlinear optical analogues (see, e.g.,
\cite{abdullaev07a,abdullaev07b,makris08,musslimani08a,musslimani08b,
  abdullaev10,bludov10,makris11,abdullaev11}). 
But to the 
best of our knowledge the basic problem of two $\cal PT$ symmetric
delta-function double wells with Gross-Pitaevskii nonlinearity
has not been considered so far.

The Gross-Pitaevskii equation we analyze in this paper has the
form
\begin{eqnarray}\label{GPEdimless}
  -\Psi^{\prime\prime}(x) &-&\left[(1+i\gamma)\delta(x+b) + (1-i\gamma)\delta(x-b)\right]\Psi(x) \nonumber \\
  &-& g|\Psi(x)|^2 \Psi(x)
  = -\kappa^2 \Psi(x)\, ,
\end{eqnarray}
with $\kappa \in {\mathbb {C}}, \; {\rm Re}(\kappa) > 0$, and $\gamma$ real. 

It consists of two delta-function traps with distance $a$, located
at $b= \pm a/2$, with a real attractive part of the potential 
and imaginary gain/loss terms whose strengths are determined
by the parameter $\gamma$, and a nonlinear
term with amplitude $g$, which arises from the contact interaction
of the condensate atoms. Units have been chosen in such a way that the strength
of the real part  of the delta-function potential is normalized to unity.
While the delta-function
potentials are $\cal PT$ symmetric, it is not clear a priori that 
the equation itself is $\cal PT$ symmetric since this requires
the nonlinear term to be a symmetric function.

For vanishing nonlinearity, we find that the 
simple quantum mechanics model captures, for both 
eigenvalues and wave functions, all the effects
of a $\cal PT$ symmetric wave guide configuration in optics.
This is essentially due to the fact that two attractive
delta-function potentials have exactly two bound states
which correspond to the two supermodes in the wave guide arrangement.

For nonvanishing nonlinearity we have solved the Gross-Pitaevskii equation  
(\ref{GPEdimless}) numerically using a procedure in which 
the energy eigenvalues are found by a five-dimensional numerical root search. 
The free parameters which have to be adjusted in such a way that a
physically meaningful wave function is obtained are the eigenvalue $\kappa$ 
as well as initial conditions for the wave
function and its derivative. Since the overall phase is arbitrary we can
choose it such that $\Psi(0)$ is a real number. Therefore five 
real
parameters remain, viz. the real part of $\Psi(0)$, 
and the real and imaginary parts of both $\Psi^\prime(0)$
and $\kappa$. Physically relevant wave functions must be square integrable and
normalized. The normalization is important since the
Gross-Pitaevskii equation is nonlinear and the norm influences the
Hamiltonian. This gives in total five conditions which have to be fulfilled:
The real and imaginary parts of $\Psi$ must vanish for $x \to \pm \infty$, and
the norm of the wave function must fulfill $||\psi|| - 1 = 0$.

Outside the delta-function traps the Gross-Pitaevskii equation 
(\ref{GPEdimless}) coincides with the free nonlinear Schr\"odinger 
equation, which has well known real
solutions in terms of Jacobi elliptic functions 
(cf., e.g., \cite{carr00,carr01,witthaut08}). 
The function which solves the equation in the ranges $|x| > b$ for 
the attractive nonlinearity considered here and decays to zero 
for $|x| \to \infty$ is ${\rm cn}(\kappa x,1) =  1/{\rm cosh}(\kappa x)$.
We find that once the correct eigenvalues and eigenfunctions are obtained 
our numerical wave functions exactly show this
behaviour. 

\begin{figure}[tb]
  \includegraphics[width=\columnwidth]{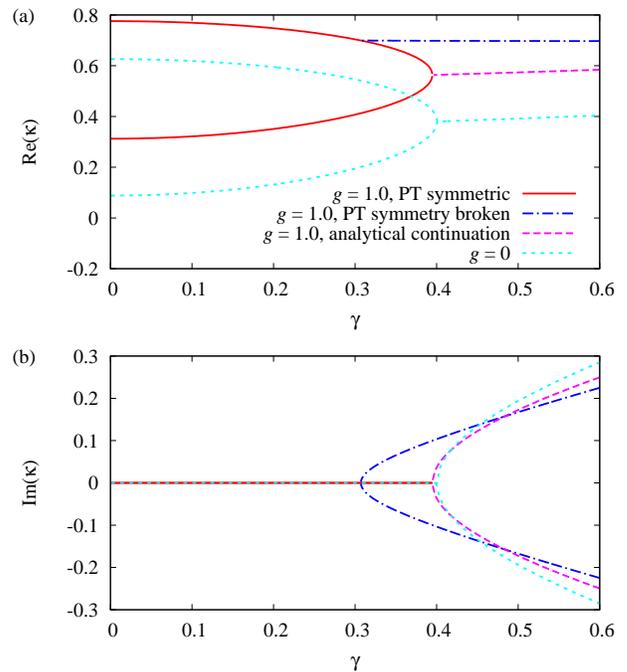}
  \caption{\label{fig:3}(Color online)
    Eigenvalues $\kappa$ of the nonlinear equation (\ref{GPEdimless})  
    as functions of the size of the loss/gain parameter $\gamma$ for $a=2.2$.
    The value of the nonlinearity parameter
    chosen is $g = 1$. The case $g = 0$ is drawn (dashed lines) for comparison. 
    A branch of complex conjugate eigenvalues appears which bifurcates
    from the ground state branch before the critical value of $\gamma$ 
    where the branches of the real eigenvalues coaelesce in an exceptional
    point. There a pair of complex eigenvalues only emerges after an analytical 
    continuation of the nonlinearity in the Gross-Pitaevskii equation.}
\end{figure}

Fig.~\ref{fig:3}
shows the results for the eigenvalues $\kappa$ calculated
for a value of the nonlinearity parameter $g=1.0$ and a trap distance of $a =2.2$  as functions of $\gamma$. The results for the case $g=0$ are
also shown for comparison. It can be seen that even with nonlinearity
there still exist two branches of real eigenvalues up to a critical value 
$\gamma_{\rm cr} \approx 0.4$, at which the two eigenvalues coincide. There also appears
a branch of two complex conjugate eigenvalues, but surprisingly these
are born, not at $\gamma_{\rm cr}$, but at the smaller value of $\gamma \approx
0.31$ where they bifurcate from the real eigenvalue branch of the ground
state. This implies that there is a range of $\gamma$ values where
two real and two complex eigenvalues coexist. 

At this point it is useful to establish a link with the
model of a $\cal PT$ symmetric Bose-Hubbard dimer with loss and gain 
investigated by Graefe et al. \cite{graefe10}.
An eigenenergy spectrum with a structure similar to the one
in Fig.~\ref{fig:3} also 
appeared in their calculations (see Fig.~13 in \cite{graefe10}).
In the model, stationary states correspond to fixed points of the 
motion of a vector on the surface of the Bloch sphere, 
whose types can be classified according to the eigenvalues of the Jacobian 
matrix.
In the region where only two real eigenvalues exist the solutions 
correspond to centers, while in the region with four 
eigenvalues the solutions correspond to a center and a saddle point,
and a sink and a source.
The center and saddle point collide at the branch point and vanish.
This behaviour is in complete agreement with the results shown 
in Fig.~\ref{fig:3}. It may be concluded that 
the familiar branching scheme known for $\cal PT$ symmetric
Hamiltonians quite generally will be changed into a scheme of the 
type as shown in Fig.~\ref{fig:3} if a nonlinearity is added to the 
Hamiltonian. 

\begin{figure}[tb]
  \includegraphics[width=\columnwidth]{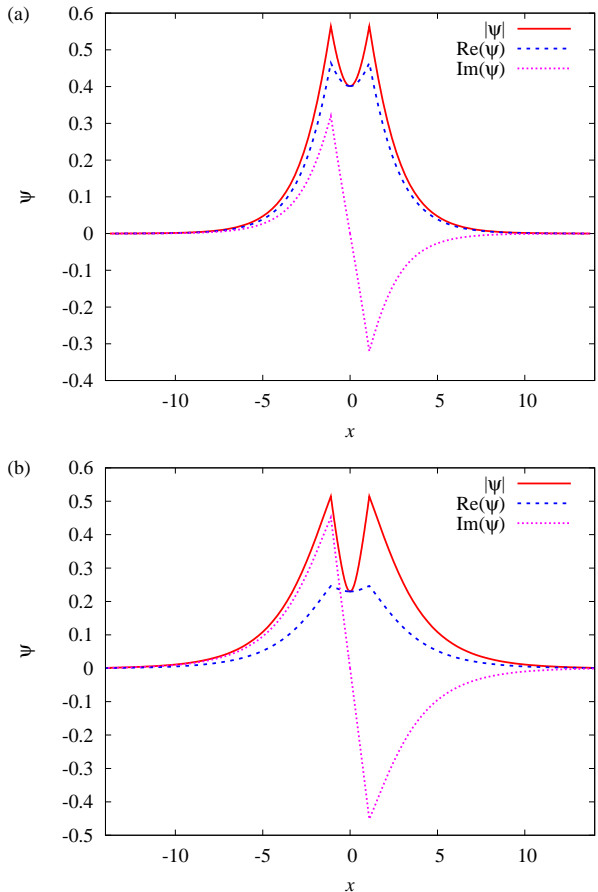}
  \caption{\label{fig:4a}(Color online)
    Real and imaginary parts and moduli of the eigenstates of the nonlinear
    Hamiltonian in equation (1) for $g=1$, $a=2.2$, 
    and $\gamma = 0.35$, (a) ground state and (b) excited state.
    The  wave functions are $\cal PT$ symmetric, and the moduli are symmetric
    functions, producing the $\cal PT$ symmetry of the total nonlinear
    Hamiltonian.}
\end{figure}

Fig.~\ref{fig:4a}
shows the real and imaginary parts of the ground state
and the excited state determined numerically for $g=1$, $a=2.2$, 
and $\gamma = 0.35$, below the the critical value 
$\gamma_{\rm cr}\approx 0.4$.
The $\cal PT$ symmetry of each wave function is evident
since their real parts are even functions and their imaginary parts odd functions
of $x$. From the $\cal PT$ symmetry of the wave function follows that the 
modulus, also shown in Fig.~\ref{fig:4a} is an even function, 
and with it
the nonlinear term in equation (\ref{GPEdimless}). We therefore have the 
important result 
that the nonlinear Hamiltonian picks as eigenfunctions exactly those states
in Hilbert space which render the nonlinear Hamiltonian $\cal PT$ symmetric!
In the ground state, which emerges from the symmetric real wave function
for $\gamma=g=0$, the symmetric contribution from the real part still
dominates, while for the excited state, which originates from
the antisymmetric solution for $\gamma=g=0$, the antisymmetric
contribution from the imaginary part prevails.

\begin{figure}[tb]
  \includegraphics[width=\columnwidth]{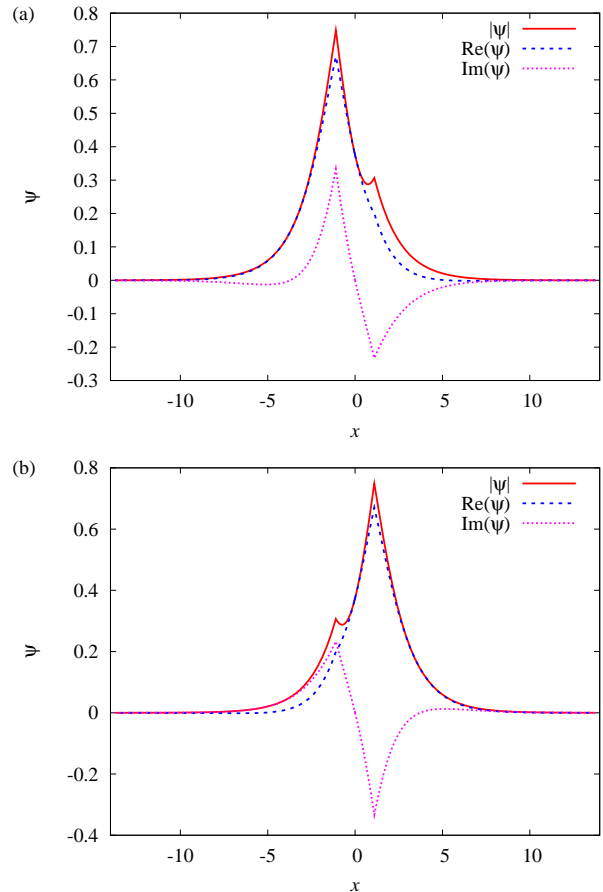}
  \caption{\label{fig:4b}(Color online) Real and imaginary parts and moduli of
    the eigenstates of the nonlinear Hamiltonian in equation (1) for $g=1$,
    $a=2.2$, and $\gamma = 0.5$, (a) solution with imaginary part of
    $\kappa > 0$ and (b) imaginary part of $\kappa <0$ . The $\cal PT$
    symmetry is broken, the moduli are not symmetric functions, and the
    $\cal PT$ symmetry also of the nonlinear Hamiltonian is broken.}
\end{figure}

The $\cal PT$ symmetry of the wave functions is broken for the eigenstates
with complex eigenvalues.
Fig.~\ref{fig:4b}
shows as an example the wave functions obtained for  $g=1$, 
$a=2.2$, and $\gamma = 0.5$ for the corresponding pair of complex conjugate 
eigenvalues $\kappa$. It can be seen that the 
real and imaginary parts are no longer even or odd functions, and therefore 
$\cal  PT$ symmetry is lost. As a consequence, the moduli of the wave functions
also  are 
no longer even functions of $x$. Thus we find that beyond the branch
point not only the $\cal PT$ symmetry of the wave functions is broken but also
that of the nonlinear Hamiltonian! 

For the states with complex eigenvalues
there is, however, an important difference between the case with and 
without nonlinearity: for complex eigenvalues the modulus squared of the wave functions grows or decays proportional to $\exp(-2{\rm Im}(\kappa^2) t)$, and so does
the nonlinear term in equation (1). Therefore strictly speaking the solutions
presented here can only describe the {\em onset} of the temporal evolution of the
two modes, for times ${\rm Im}(\kappa^2) t \ll 1$. To obtain the
full time evolution in the region beyond the branch point one would have to 
solve the time-dependent Gross-Pitaevskii equation. 
This clearly goes beyond the scope of our simple model 
calculation. In Fig.~\ref{fig:4b} the mode with positive imaginary part
of $\kappa$ is the one which begins to decay, as expected it is more 
strongly localized 
in the trap with loss, while the mode with negative imaginary part is the one
which starts to grow and is more strongly localized in the trap with gain.

The fact that at the branch point two real solutions coalesce 
without giving rise to two solutions with complex eigenvalues
contradicts the usual behaviour seen at exceptional points. 
Obviously these solutions cannot be found by solving the
nonlinear Gross-Pitaevskii equation in its form (\ref{GPEdimless}),
but require an analytical continuation of the nonlinear Hamiltonian
beyond the critical point $\gamma_{\rm cr}$. The reason is that the
nonlinear term $g|\Psi|^2$ is a nonanalytic function, and some
care has to be taken when analytically continuing the Hamilitonian
beyond the exceptional point.

In the $\cal PT$ symmetric regime up to $\gamma_{\rm cr}$ 
we have $\Psi^\ast(x) = \Psi(-x)$. Therefore
on the way to the bifurcation point we can write the nonlinearity
for the $\cal PT$ symmetric states in the form 
$g|\Psi(x)|^2 \equiv g \Psi(x) \Psi(-x)$. 
This function can be continued analytically. In the numerical calculation
the additional condition $\int \Psi(x) \Psi(-x) dx = 1$ must be introduced
to fix the phase of the nonlinearity in the $\cal PT$ broken
regime. In the (then) six-dimensional root search also ${\rm Im}(\Psi(0))$ must
varied. As a result we find two more complex conjugate solutions 
that emerge from the coaelescing states, see Fig.~\ref{fig:3}.
These states are not $\cal PT$ symmetric, and no longer possess vanishing imaginary parts at the origin. 

In this paper we have analyzed the simple quantum mechanical 
model of a Bose-Einstein condensate
in $\cal PT$ symmetric delta-function double traps by directly solving the 
nonlinear Gross-Pitaevskii equation. 
We find two stationary eigenstates with real eigenvalues
which  at a critical value of the 
loss/gain parameter merge in a branch point.
We have the important result that the wave functions are
$\cal PT$ symmetric. 
As a consequence their moduli are even functions,
and therefore the nonlinear Hamiltonian selects as solutions exactly
such states which make itself $\cal PT$ symmetric. We also find a 
branch of two complex conjugate eigenvalues for which   
the $\cal PT$ symmetry of the wave functions is broken, and 
with it that of the nonlinear Hamiltonian. 

An unexpected result
is that, with the nonlinearity present, the branches of complex conjugate
eigenvalues do not bifurcate from the point where the real
eigenvalues conincide, but emerge at a smaller value of the gain/loss 
term from the eigenvalue branch of the ground state. 
On the other hand, at the critical value of the gain/loss parameter
we find the behaviour characteristic of a branch point, i.e.
the coalescence of both eigenvalues and eigenfunctions, but no
pair of complex conjugate eigenvalues seems to emerge. These
are found only after continuing analytically the nonlinear term 
in the Gross-Pitaevskii equation. 
Note, however, that as stated before,
for complex eigenvalues  the squared modulus of the wave function
becomes time-dependent, and a description using the stationary
Gross-Pitaevskii equation breaks down anyway. 
This does, however, not affect the main result of our paper, namely
the existence of $\cal PT$ symmetric eigenfunctions and the 
$\cal PT$ symmetry of the Hamiltonian also when the nonlinearity
is present.

We have considered the 
case of an attractive nonlinearity 
but found that the same behaviour occurs for repulsive nonlinearity.

The results of our model calculation make one to  expect that similar
$\cal PT$ behaviour should also prevail in Bose-Einstein 
condensates in more realistic double wells \cite{ananikian06} with 
$\cal PT$ symmetry, in one or more dimensions. Also in addition
to the nonlinearity resulting from the short-range contact interaction
condensates with a long-range dipole-dipole interaction \cite{lahaye09b}
could be considered. Investigations of the Gross-Pitaevskii equation in these 
directions are under way. It would also be interesting to 
extend the quasi-Hermitian analysis given by Jones \cite{jones08}
and to investigate whether
for the nonlinear $\cal PT$ symmetric Hamiltonian considered in the present
paper the construction of a metric operator is possible with respect to
which the nonlinear Hamiltonian is quasi-Hermitian. 
Furthermore it would be worthwhile looking for simple matrix models
which show the behaviour of the eigenvalues found for finite nonlinearity.

\begin{acknowledgments}
  We thank Eva-Maria Graefe and Miloslav Znojil for helpful comments.
\end{acknowledgments}

\end{document}